# Fabrication Process and Properties of Fully-Planarized Deep-Submicron Nb/Al-AlO$_x$/Nb Josephson Junctions for VLSI Circuits

Sergey K. Tolpygo, Vladimir Bolkhovsky, Terence J. Weir, Leonard M. Johnson, *Senior Member, IEEE*, Mark A. Gouker, *Senior Member, IEEE*, and William D. Oliver, *Member, IEEE*

*Abstract*—**A fabrication process for Nb/Al-AlO$_x$/Nb Josephson junctions (JJs) with sizes down to 200 nm has been developed on a 200-mm-wafer tool set typical for CMOS foundry. This process is the core of several nodes of a roadmap for fully-planarized fabrication processes for superconductor integrated circuits with 4, 8, and 10 niobium layers developed at MIT Lincoln Laboratory. The process utilizes 248 nm photolithography, anodization, high-density plasma etching, and chemical mechanical polishing (CMP) for planarization of SiO$_2$ interlayer dielectric. JJ electric properties and statistics such as on-chip and wafer spreads of critical current, $I_c$, normal-state conductance, $G_N$, and run-to-run reproducibility have been measured on 200-mm wafers over a broad range of JJ diameters from 200 nm to 1500 nm and critical current densities, $J_c$, from 10 kA/cm$^2$ to 50 kA/cm$^2$ where the JJs become self-shunted. Diffraction-limited photolithography of JJs is discussed. A relationship between JJ mask size, JJ size on wafer, and the minimum printable size for coherent and partially coherent illumination has been worked out. The $G_N$ and $I_c$ spreads obtained have been found to be mainly caused by variations of the JJ areas and agree with the model accounting for an enhancement of mask errors near the diffraction-limited minimum printable size of JJs. $I_c$ and $G_N$ spreads from 0.8% to 3% have been obtained for JJs with sizes from 1500 nm down to 500 nm. The spreads increase to about 8% for 200-nm JJs. Prospects for circuit densities $> 10^6$ JJ/cm$^2$ and 193-nm photolithography for JJ definition are discussed.**

## I. INTRODUCTION

SINGLE-FLUX-QUANTUM (SFQ) superconductor electronics (SCE) has well documented advantages over semiconductor electronics for energy efficiency, clock speed, and potential for reversible computing. Over the years, however, these advantages have largely remained unrealized because of low scale of integration and limited sophistication in functionality of superconducting circuits. While the semiconductor industry maintained an exponential growth of electronic circuit density, the density of JJ circuits has remained nearly stagnant for the last 25 years. This stagnation has been attributed in part to the gross disparity in funding of SCE compared with CMOS and the resultant lack of access to modern fabrication tools and systematic process development [1]-[3].

Achieving very large scale integration (VLSI) with $10^6$ JJ/circuit and beyond requires decreasing the scale of all components to deep submicron dimensions. While there have been a number of reports on submicron JJs [4]-[11] with both low and high critical current density, $J_c$, the JJ definition was produced with e-beam lithography in circuits with a small number of JJs on small-size wafers [12]. There are no data and statistics on submicron JJs defined by modern steppers and scanners with 248 nm and 193 nm exposure wavelength, on a very large scale and on large-size wafers.

The goal of this work has been to develop a robust fabrication process for Nb/Al-AlO$_x$/Nb junctions, using 200-mm production class tools of a typical CMOS foundry. In this paper we report on JJ fabrication as well as yield, JJ quality, repeatability and uniformity statistics on a large set of JJs in order to assess the JJ technology for VLSI circuits. This work builds on a decade of fully planar, niobium-trilayer process development targeting superconducting qubits for quantum information science and technology applications. In the present context, however, we focus solely on aspects related to the development of a robust niobium JJ fabrication process for classical digital and mixed signal applications. Further, the broader effort includes increasing the number of niobium wiring layers from 8 in the current process [13] to 10, and decreasing the size of inductors [14], resistors, and vias [15] to achieve circuits densities > 1M JJs/cm$^2$.

## II. JOSEPHSON JUNCTION FABRICATION PROCESS

### A. General process description

High-resolution photolithography requires high planarity of circuit layers because of the small focus depth $\sim \lambda/(NA)^2$ of modern photolithography tools using a short exposure wavelength, $\lambda$, and a high numerical aperture, $NA$, objective lenses. Therefore, in the MIT LL fabrication process all layers are planarized using chemical mechanical polishing (CMP) of dielectric layers. This ensures all metal layers are deposited on a nearly flat surface with topography height less than ~ 40 nm, except near etched vias in the dielectric which provide interlayer connections. A planarized stud-via process has also been developed [15] but is not used in this work.

Planarization also allows the junction layer to be placed at any process level, since the layers are virtually identical, such that the choice may be dictated by SFQ circuit design considerations. For example, in our 4-metal-layer (4M) process, JJs are placed over one Nb ground plane layer, whereas in the 8-metal-layer (8M) process the JJ trilayer is above 4 planarized layers of Nb wiring, as shown in Fig. 1.

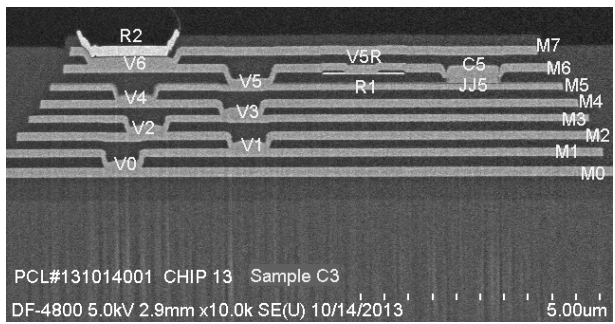

Fig.1 Cross-section of MIT LL 8-metal-layer fabrication process with 8 fully-planarized Nb layers marked as M0 – M7. JJ is marked JJ5, resistor R1, contact pad R2. Etched vias between adjacent layers are marked V0, V1, etc.

Nb/Al-AlO$_x$/Nb trilayers were deposited in-situ in a cluster tool with separate chambers for magnetron sputtering of Nb, Al, and for AlO$_x$ formation by thermal oxidation. Nb base and counter electrodes were 150 nm and 250 nm, respectively, and the Al layer was 8-nm thick. The trilayer was deposited over a 200-nm thick planarized SiO$_2$ layer covering the bottom wiring layers M0-M4 in the 8M process or just one layer, M4, of the 4M process. 200-mm oxidized Si wafers were used with, typically, eight wafers per lot. Oxygen exposure (pressure·time) for oxidation was adjusted to target the following Josephson critical current densities: 10 kA/cm$^2$, 20 kA/cm$^2$, and 50 kA/cm$^2$ [16]. Most of the data presented hereafter were obtained using our main, 10-kA/cm$^2$ process, from hundreds of processed wafers. The data for higher $J_c$ junctions were obtained on a much smaller set of wafers used to explore these $J_c$ values as the targets for future process nodes.

### B. Photolithography of submicron Josephson junctions

We used a Canon FPA-3000 EX4 stepper with 248 nm KrF excimer laser exposure wavelength, 5x reduction and $NA$=0.6 for photolithography of all layers. We used a positive photoresist and antireflection coating to minimize standing waves caused by the high reflectance of Nb layers.

The optical patterning of JJs reduces to photolithography of posts or contact holes in semiconductor manufacturing. The important differences are: first, a high accuracy and precision of the area definition is required as the critical current $I_c$ is proportional to JJ area, and second, a broad range of JJ sizes needs to be printed in a manner preserving the proper relationship between their areas and critical currents.

We used a circular-shaped design for JJs, as it retains circular symmetry after diffraction and gives the minimum area for a given critical current (CD). The square shape maximizing the area and used in [17]-[19] requires resolution enhancement techniques (RET) to correctly print corners rounded by diffraction effects, making the area control more difficult. In practice, the JJs were circular-shaped polygons (manhattans) defined on a 5-nm grid (25-nm grid on 5x photomask). Hereafter we give all dimensions in the 1x wafer scale after the 5x reduction.

The theoretical resolution limit for a line-space structure with pitch $p = w + s$ is $p_{min} = (\lambda/NA)\cdot 1/(1+\sigma)$, where $w$, $s$, and $\sigma$ are the linewidth, space, and exposure coherence factor, respectively [20]. This gives $p_{min}$ = 413 nm for coherent exposure ($\sigma = 0$) and 250 nm for partially coherent exposure with $\sigma = 0.65$. There is no theoretical resolution limit for single circles or squares. A practical limit exists, however, and it is set by the photoresist properties, focus depth budget, and the desired linearity of the relationship between the printed size on the wafer, $d_w$, and the drawn size on the mask, $d$.

Typical SEM images of photoresist features used for JJ definition are shown in Fig. 2. After photoresist exposure and development, the diameter of obtained features, i.e., the photoresist mask for JJ etching, was measured using a Hitachi CD-SEM by taking the mean of the measurements of the feature diameter at 49 points around the feature perimeter, as shown by the dotted white line in the right panel of Fig. 2. The measured dependence of the photoresist features on the wafer, $d_w$, on the drawn size, $d$, is shown in Fig. 3. It is similar to the dependence we reported earlier in [15] for printing Nb posts in the context of our stud-via technology. Over a very wide range of JJ diameters, the dependence is very well approximated by

$$d_w = (d^2 - d_c^2)^{\frac{1}{2}} + b \qquad (1)$$

for $d > d_c$, and $d_w = 0$ otherwise, where $d_c$ is the minimum printable size (i.e., the photolithography cut-off) and $b$ is the process bias. The typical value of $d_c$ we observed over a several hundred wafers is 245 nm ± 20 nm, and the minimum value was ≈ 225 nm.

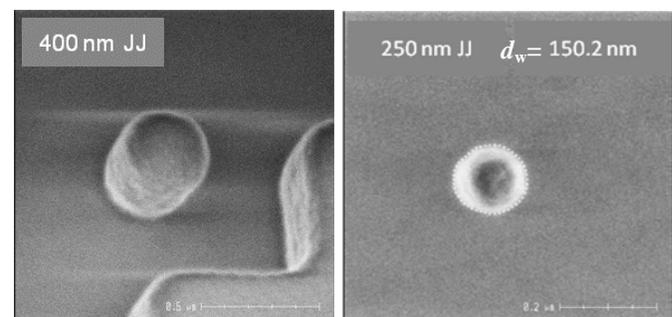

Fig. 2. SEM images of photoresist features. Left panel shows a tilted view of the photoresist JJ etch mask at $d$ = 400 nm. Right panel shows the top view of the photoresist JJ etch mask at $d$ = 250 nm, giving $d_w$ = 150 nm; white dots around the perimeter show 49 points used for measuring the mean feature diameter, $d_w$.

The origin of the cut-off is as follows. The requirement to print a range of JJ sizes sets an exposure dose - dose to print -

that gives the proper sizing of the desired features. Diffraction of light on the mask features creates a nonzero exposure of the photoresist in the "dark" region of the geometric shade. This unwanted exposure increases as the feature size decreases, Fig. 4. At a certain size, $d_c$, the exposure exceeds a critical level at any point on the wafer and renders all the photoresist completely soluble in the developer; the photoresist mask develops away in the development process.

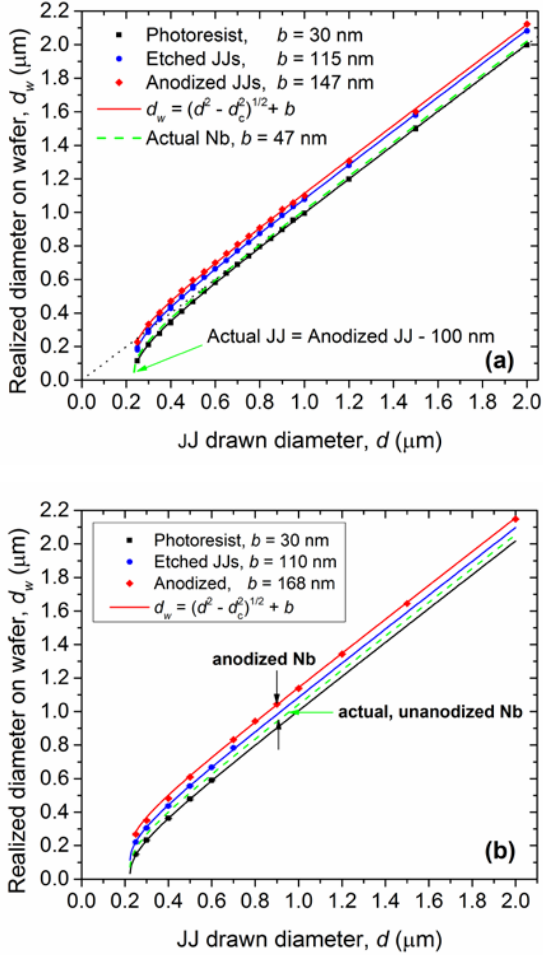

Fig. 3. The on-wafer diameter, $d_w$, of circular features formed during the steps of Josephson junction fabrication as a function of JJ drawn diameter on the photomask. The data points with fitted curves from the bottom to top in the figures correspond to: (■) – photoresist etch mask, (●) - etched counter electrode of JJs, (♦) - anodized counter electrode of JJs. Solid curves show fits to (1), giving $d_c$ = 235 nm and 223 nm, respectively, for the data in Fig. 3(a) and Fig. 3(b). The data in Fig. 3(a) and Fig. 3(b) were taken during two fabrication runs of the 8-metal layer process, about a year apart, and demonstrate the high stability and reproducibility of the fabrication process. The size bias increases from $b$ = 30 nm for the photoresist etch mask to ~ 150 nm after anodization of JJs. The real diameter of the un-anodized Nb counter electrode, $i.e.$, the actual JJ size, is shown by dashed curves. It is ≈ 100 nm less than the anodized JJ seen in the SEM. The desired, ideal, relation $d_w = d$ is shown by the dotted line in Fig. 3(a).

In order to provide a mathematical description of the image size for comparison with the measurements, we use the standard theory of image formation in optical imaging systems based on Fraunhofer scalar diffraction theory and Fourier optics [21],[22]. In this theory, the image formed by an objective is a Fourier transform of the diffraction pattern of the optical mask filtered by the finite numerical aperture of the objective (exit pupil). For coherent illumination, the amplitude of light in the image plane on a wafer can be written as

$$E(x,y) = \iint_{-\infty}^{+\infty} \tilde{P}(f,g)\tilde{O}(f,g)e^{-i2\pi(fx+gy)}dfdg, \quad (2)$$

where $\tilde{O}(f,g)$ is the mask spectrum, $i.e.$, the Fourier transform of the mask pattern, $\tilde{P}(f,g)$ is the so-called pupil function describing the response of the projection system to spatial frequencies $(f,g)$, and the term $\exp[i2\pi(fx+gy)]$ represents the interference of light rays propagating in a direction $\theta$ given by $|\sin\theta| = \lambda\sqrt{f^2+g^2}$ [23].

For an opaque disk with diameter $d$ on the photomask (a drawn JJ), the mask transmittance function is

$$O(x,y) = 1 - circ(2r/d)$$
$$O(x,y) = 0 \text{ if } 2r/d \leq 1, \text{ and } O(x,y) = 1 \text{ if } 2r/d > 1 \quad (3)$$

where $r = \sqrt{x^2+y^2}$. The Fourier transform of the $circ(r)$ function is an Airy function [22], so the mask spectrum is

$$\tilde{O}(f,g) = \delta(f)\delta(g) - \frac{\pi d^2}{2}\frac{J_1(\pi dq)}{(\pi dq)}, \quad (4)$$

where $q = \sqrt{f^2+g^2}$, $\delta$ is Dirac delta-function, and $J_1(u)$ is a first-order Bessel function of the first kind.

For a circular-symmetric imaging system with coherent illumination, the pupil function is

$$\tilde{P}(f,g) = 1 \text{ if } \sqrt{f^2+g^2} \leq NA/\lambda, \text{ and 0 otherwise.} \quad (5)$$

That is, an image is formed only by the light rays arriving at angles with $|\sin\theta|$ less than the numerical aperture of the lens, and all higher spatial frequencies are filtered out. This sets the limits of integration in (2) from $-NA/\lambda$ to $+NA/\lambda$.

In general, (2) requires numerical integration. In order to get an analytical expression, we consider the case of very small mask diameters $d \ll \lambda/NA$ and use a zeroth-order expansion of the Airy function $J_1(u)/u = 1/2$. Then, the mask function in (4) becomes

$$\tilde{O}(f,g) \approx \delta(f)\delta(g) - \pi d^2/4, \quad (6)$$

and the integral in (2) can be evaluated [22] to obtain the field distribution in the image plane, $i.e.$ the aerial image,

$$E(\rho) \approx 1 - \frac{1}{4}\left(\frac{\pi d NA}{\lambda}\right)^2 \frac{2J_1\left(\frac{2\pi\rho NA}{\lambda}\right)}{2\pi\rho\left(\frac{NA}{\lambda}\right)}, \quad (7)$$

which retains the circular symmetry of the photomask, and $\rho$ is the distance from the center of the aerial image. The normalized intensity of light at any point is $I(\rho) = E^2(\rho)$.

To find the size of the image realized in the photoresist, we apply the standard model of an infinitely thin, threshold resist [23], [24]. In this model, the photoresist is developed away at any point on the wafer that receives an exposure dose higher than a threshold dose corresponding to the normalized light intensity $I_{th}$ and remains on the wafer otherwise. The contour of the image in the image plane is then given by points satisfying the solution of the equation

$$E(x, y)^2 = I_{th},$$  (8)

which reduces to $E(\rho) = \sqrt{I_{th}}$ in our circular-symmetric case. A solution of this equation exists only if the mask diameter (JJ drawn size) is larger than the minimum size $d_c$, given by

$$d_c = \frac{2\lambda}{\pi NA} (1 - \sqrt{I_{th}})^{1/2}.$$  (9)

Otherwise, the light intensity at any point of the image is larger than $I_{th}$ and the resist will be developed away. Indeed, at the center of the image where $2J_1(u)/u = 1$, the light intensity is $\left[1 - \left(\frac{\pi dNA}{2\lambda}\right)^2\right]^2$, and (8) can only be satisfied if $d \geq d_c$ as given by (9). A similar treatment for square posts can be found in [25], giving for the minimum printable size of squares an extra factor of $2/\sqrt{\pi}$ in (9).

For mask diameters slightly larger than $d_c$, the size of the image in photoresist can be found by expanding the Airy function in (7) to the next leading order, $2J_1(u)/u = 1 - u^2/8$ to obtain, using (8),

$$d_w = \frac{2\sqrt{2}\lambda}{\pi dNA} (d^2 - d_c^2)^{1/2},$$  (10)

with $d_c$ given by (9). For $d - d_c \ll d_c$, (10) gives the same functional dependence as the empirical relation (1). Note, that according to (7) the light intensity grows as $d^4$ for $d \ll NA/\lambda$, so the cut-off at $d_c$ is very sharp.

An intensity threshold $I_{th} = 0.3$ is a good representation of the typical photoresists [23]–[25]. This gives $d_c = 177$ nm for our exposure conditions, a bit lower than the typical values of $d_c > 220$ nm observed in this work. The difference is attributable to a partially coherent exposure used, finite photoresist thickness, defocusing, and other photoresist effects.

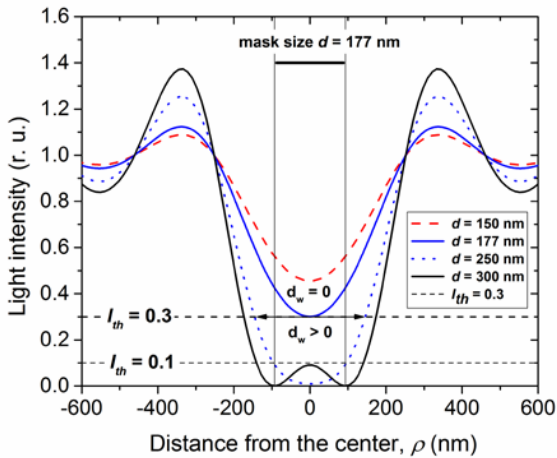

Fig. 4. Light intensity distribution in the image plane (aerial image of JJs) for several diameters of the opaque disc representing various drawn JJs on the photomask. The distance is counted from the center of the aerial image. If $d \leq d_c$ (e.g., two upper curves), the light intensity everywhere is larger than the photoresist exposure threshold $I_{th}$ so that there is no image photoresist, i.e., all the resist is developed away, and hence there is no photoresist mask for JJ etching. The diameter of the resultant photoresist image as well as the cut-off diameter $d_c$ depends on the photoresist threshold intensity.

Light intensity distributions following from (7) are shown in Fig. 4 for several mask diameters $d$ below and above the minimum printable size $d_c$. A photoresist with lower $I_{th}$, e.g., $I_{th} = 0.1$ also shown in Fig. 4, gives a larger value of $d_c$ and a smaller photoresist image.

For a wider range of sizes, approximation (6) for the mask function (4) is insufficient and (2) needs to be integrated numerically. Numeric results obtained using the photolithography simulator PROLITH 9.0 [26] are shown in Fig. 5 for coherent and partially coherent illumination. Not surprisingly, near the cut-off size the numerical results agree very well with our analytical result (10) for coherent illumination. For the partially coherent illumination, $\sigma = 0.65$, the cut-off diameter increases to 192 nm, closer to the $d_c$ observed.

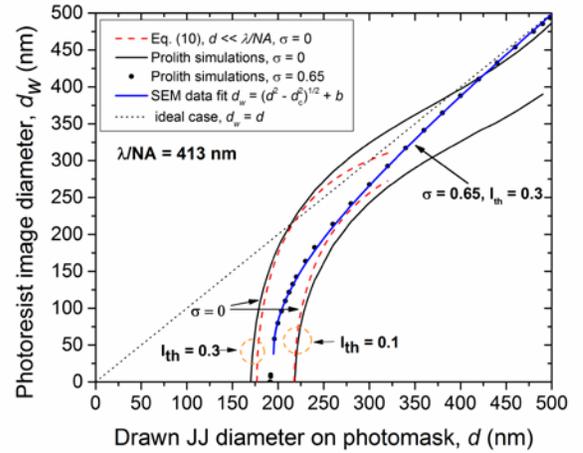

Fig. 5. Results of the analytic and numeric calculations of the image diameter in the threshold photoresist as a function of the drawn diameter on the photomask for different exposure parameters and photoresist threshold intensity, $I_{th}$. Dashed lines are the analytical result (10) for coherent illumination, $\sigma = 0$, and $d \ll NA/\lambda$ at two different threshold intensities; solid lines are numeric results obtained using PROLITH for the same parameters. PROLITH simulations at $\sigma = 0.65$ are shown by dots (●) along with the fit to (1) shown by the solid line; they are virtually indistinguishable.

The simulated photoresist image diameter at $\sigma = 0.65$ (used in this work) is very well fitted by the empirical function (1) that was derived to fit the photoresist feature diameters measured in the SEM, see Fig. 3 and Fig. 5. This gives yet another justification for using (1) in further discussions of JJ properties.

## C. Etching and anodization of Josephson junctions

After patterning the photoresist mask, the JJ counter electrode was etched using a high-density plasma etcher and then anodized to prevent degradation of the exposed AlO$_x$ barrier [27]-[29]. The sizes of the etched JJs were measured before and after the anodization step, and are shown in Fig. 3 along with the size of the photoresist features used as an etch mask. For statistics, multiple JJs of the same nominal size were measured on the same test chip at different locations on the wafer.

We developed an etch recipe that provides a constant positive bias $\Delta b \approx 80$ nm to the etched features, thereby obtaining a larger diameter of the etched JJs than the photomask diameter in order to compensate for a consumption of Nb in the anodization process. The post-etch diameter of JJs is shown in Fig. 3 by blue dots along with the fits to (1), shown by blue curves. The post-etch diameters differ from the etch-mask diameters, shown in black, by a size-independent bias – *i.e.*, all the data points are shifted up by the same amount.

Anodization adds yet another bias component, increasing the JJ diameter by about 50 nm, Fig.3. This increase is caused by the $Nb_2O_5$ layer grown on the sidewalls of JJs during the anodization. Its thickness was measured using cross-section TEM, and was found to be $\approx 50$ nm, see Fig. 6.

During the anodization, about 0.9 nm/V of Nb is consumed and converted in about 2.5 nm/V of niobium oxide. Therefore, the actual diameter of niobium in the counter electrode of JJs, *i.e.*, the final JJ size as seen in SEM, is given by $d_w = d_{anod} - 100$ nm , where $d_{anod}$ is the JJ diameter after the anodization. This dependence is shown in Fig. 3(a) and (b) by a green dashed curve; it corresponds to the final bias $b$ in the range from 50 nm to 70 nm in (1).

The final bias is the sum of the photolithography, etching, and anodization components: $b = b_{photo} + b_{etch} + b_{anod} - 100$ nm. The purpose of the described bias settings was to obtain the final diameter of Nb in the counter electrode of JJs within $\pm 50$ nm from the diameter of the photoresist etch mask. This is convenient from the in-line process control standpoint. Note that etching and anodization do not affect $d_c$, which is determined solely by the photolithography process.

### D. Planarization and further processing of JJs

After patterning the base electrode of the JJs using the photolithography and high-density plasma etching, the junction mesa was planarized by depositing a thick layer of $SiO_2$ and polishing it down to the level of the JJ counter electrode by using CMP. This creates a flat surface of the interlayer dielectric, on to which a Nb wiring layer is deposited and patterned for contacting the tops of the JJs. The $SiO_2$ thickness was measured using an ellipsometer. The surface roughness of $SiO_2$ layers after the CMP was measured by an AFM and found to be less than 0.4 nm over a 10 µm x 10 µm window.

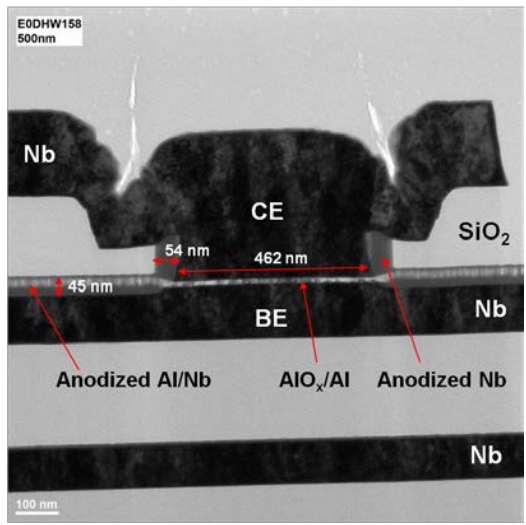

Fig. 6. A cross-section TEM image of fully processed Josephson junction with drawn diameter $d = 500$ nm. Lighter region along JJ sidewalls and base electrode is anodized niobium and Al/Nb bilayer of the base electrode (BE).

Further processing steps involve depositing and patterning the resistor layer, adding an interlayer dielectric above the resistor layer, opening contact holes to the buried JJ base electrode and resistors, and depositing and patterning the wiring layer used to contact the JJs and the resistors; see the final process cross-section in Fig. 1. A cross-section TEM of a fully processed 500-nm JJ is shown in Fig. 6.

### E. Electrical characterization of Josephson junctions

The electric resistance of individual JJs on the processed wafers was measured by semi-automated wafer probers, using 4-probe measurements of JJs in a cross-bridge Kelvin geometry (CBKR). For a description of CBKR technique see [15],[31],[32],[61],[62] and references therein. A JJ design with a minimum surround by the counter electrode and the top wiring layer was used to minimize its parasitic contribution of the JJ tunnel barrier resistance. For the JJ sizes used in this work, this parasitic contribution is negligible, but it may become noticeable at $d > 5$ µm and even less for higher-$J_c$ junctions. To test JJ repeatability, a number (typically 36, 44 or 60) of nominally identical JJs were measured on each of 9 dies spread across the wafer, providing data on 324 - 540 JJs for each size, and over 4000 JJs in total per wafer. JJ conductance versus size dependences were used to infer the tunnel barrier specific conductance and to estimate the $J_c$ from room-$T$ measurements, using 160 test structures with JJ drawn sizes from 200 nm to 15 µm at each location. A few test structures with JJ drawn sizes $d \leq d_c$ were used to measure $d_c$ electrically and to assess its reproducibility on the wafer scale. Junctions with $d \leq d_c$ were generally tested open, consistent with SEM measurements done on a few chips per wafer.

The test structures from the same set, but with $d > d_c$, were then measured at LHe temperature after dicing the wafers and wire bonding devices on individual chips. Current-voltage ($I$-$V$) characteristics of single JJs and 1000-JJ series arrays were measured in a magnetically shielded cryoprobe allowing for automated testing of up to fourteen 5 mm by 5 mm chips in liquid He.

### III. EXPERIMENTAL RESULTS AND DISCUSSION

### A. General approach, justification, and statistics

Our approach is to evaluate continually the reproducibility (between wafers) and repeatability (within a wafer) of the fabrication process and its components, *e.g.*, JJs, on a very large scale. Cryogenic measurements of individual JJs on this scale, however, are practically prohibitive due to expense and time investment. Also, critical currents of small junctions with $I_c$ less than about $20 - 50$ µA cannot be directly measured at 4.2 K because their switching is affected by noise. Procedures for extracting $I_c$ in such cases exist, *e.g.*, measuring the switching statistics and subsequent fitting to a theoretical dependence, or measuring the devices in a dilution refrigerator at much lower temperatures. However, these techniques even more time consuming and expensive for the task at hand. Instead, our approach is to establish a clear connection between wafer-scale statistical data (on a large number of devices per wafer) acquired at room temperatures and cryogenic measurements (fewer devices per wafer) that are used to validate and supplement them.

It has been very well established during the last 50 years that the critical current of a Josephson junction and its normal state conductance $G_N = 1/R_N$ are related, because both depend on the very same property of the junction - the barrier transparency, for a review see, *e.g.*, [47]. This is commonly expressed by the statement that $I_c R_N$ at a given temperature is a constant for the given type of JJs. For tunnel junctions the relationship is known as Ambegaokar-Baratoff (AB) formula [33]

$$I_c = G_N V_{AB}(T) \text{ and } V_{AB}(T) = g(\pi V_g/4)\tanh(V_g/4k_B T), \quad (11)$$

where $V_g$ is the gap voltage $(\Delta_1+\Delta_2)/e$, $\Delta_1(T)$ and $\Delta_2(T)$ are the superconducting energy gaps in the junction electrodes, and $g$ is a numerical factor of order unity which accounts for deviations from the ideal barrier [33], such as barrier non-ideality, proximity between Nb and Al overlayer in the base electrode, strong coupling effects in Nb, *etc.*

We determined $V_{AB}$ at 4.2 K for our junctions and used it to infer the critical current of JJs from their conductance. Also, we experimentally established the relation between the normal state conductance $G_N$ at 4.2 K, entering into the AB formula (11), and the JJ conductance $G_{300}$ at 300 K. This chain of relationships allows us to use $G_{300}$ to characterize the value of $I_c$ of the junctions at 4.2 K.

However, when considering the relative statistical quantities, *e.g.*, on-chip and on-wafer standard deviation of $I_c$ and run-to-run reproducibility, the knowledge of the exact value of $V_{AB}$ is not strictly required in all cases of interest.

For example, if quantities $G_N$ and $V_{AB}$ are independent statistical variables $X$ and $Y$, the variance of their product is $Var(XY)=Var(X){\cdot}Var(Y) + Var(X){\cdot}[E(Y)]^2+Var(Y){\cdot}[E(X)]^2$, where $E(X)$ and $E(Y)$ are the expectation values, *i.e.*, their mean values [48]. Hence, a normalized value of $I_c$ standard deviation $\sigma_{Ic} = [Var(I_c)]^{1/2}/E(I_c)$ can be written as

$$\sigma_{Ic} = (\sigma_G^2\sigma_{Vab}^2 + \sigma_G^2 + \sigma_{Vab}^2)^{1/2}, \quad (12)$$

where $\sigma_G$ and $\sigma_{Vab}$ are the normalized standard deviations of the JJ conductance and the Ambegaokar-Baratoff voltage - $I_c R_N$ product, - respectively. In turn, the $\sigma_{Vab}$ is the same as the relative standard deviation of the gap voltage $\sigma_{Vg}$ in the ensemble of junctions since the variation of the hyperbolic tangent term in (11) is negligible at 4.2 K where $V_g >> k_B T$.

Thus, the standard deviations $\sigma_G$ and $\sigma_{Vg}$ can serve as a good estimator of $\sigma_{Ic}$ by using (12), and a complex problem of the exact determination of $I_c$ and its statistics, especially in small JJs, is replaced by the simpler task of finding $\sigma_G$ and $\sigma_{Vg}$ from relatively straightforward measurements of the gap voltage and JJ conductance.

The results of the implementation of the approach described above are given in the subsequent sections and followed by a discussion.

### B. 10 kA/cm² process

Fig. 7 shows the relation between the room-temperature JJ resistance $R_{300}$ and JJ normal state resistance $R_N$ extracted from *I-V* characteristics measured at 4.2 K. The relation is linear, $R_{300} = kR_N$ with $k = 0.905\pm0.010$ and, respectively, $G_N = kG_{300}$. Values of $k$ less than unity are consistent with the tunnel barrier and indicate negligible contribution of Nb leads to the junction resistance [30]-[32]. *I-V* characteristics also show high junction quality as indicated by a relatively large subgap resistance $R_{sg}/R_N > 10$ and gap voltage $V_g = 2.75$ mV. The AB voltage $I_c R_N = V_{AB}(T)$ at 4.2 K was determined to be 1.75 mV for our 10 kA/cm² JJs.

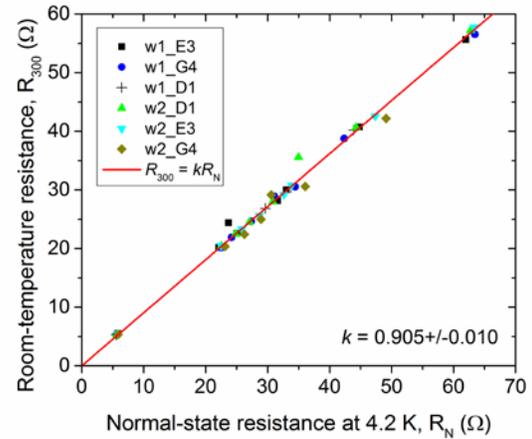

Fig. 7. The room temperature resistance of JJs in CBKR configuration *vs* their normal state tunnel resistance from *I-V* characteristics at 4.2 K for multiple junctions and different locations on two wafers of 10 kA/cm² process.

The linear relation between $R_{300}$ and $R_N$ (and respective conductances $G_{300} = 1/R_{300}$ and $G_N$) enables us to use the automated wafer-prober measurements of $R_{300}$ to screen rapidly a very large number of JJs on 200-mm wafers, in lieu of performing time consuming measurements at 4.2 K. This enables a statistical assessment of JJ uniformity and reproducibility on the wafer scale, and it provides an accurate prediction of JJ critical currents and trilayer's $J_c$ based on room-$T$ measurements.

The gap voltage standard deviation $\sigma_{Vg}$ was measured using single JJs and series arrays of 100 and 1000 JJs with drawn sizes from 500 nm to 2200 nm. It was determined to be less than 0.4%, limited by the accuracy of our voltage measurements (about $\pm10$ $\mu$V). Consequently, the first and the third term in (12) can be neglected, and the critical current standard deviation is primarily determined by the standard deviation of JJ conductance.

The JJ conductance, both $G_N$ and $G_{300}$, and the critical current $I_c$ are proportional to JJ area

$$G_{300} = G_0 \frac{\pi}{4}(d_w - d_0)^2 \text{ and } I_c = J_c \frac{\pi}{4}(d_w - d_1)^2, \quad (13)$$

where parameters $d_0$ and $d_1$ account for the deviation in the junction diameter as determined by electrical measurements at, respectively, 300 K and 4.2 K from its physical diameter $d_w$ as measured by the SEM. Since $d_w$ is described by (1), parameters $d_0$ and $d_1$ simply modify the size bias $b$. Parameters $G_0$ and $J_c$ characterizing the tunnel barrier can be found by fitting $\sqrt{G_{300}}$ and $\sqrt{I_c}$ to (13) with $d_w$ given by (1) with $G_0$, $J_c$, and $b$ taken as fitting parameters. An example of such a fitting is shown in Fig. 8 for nine locations on a 200-mm wafer. All data are fitted exceptionally well by accounting for the nonlinear relationship between the JJ electric size and the drawn size due to the optical process bias described by (1). Simple fits assuming solely a constant bias, the so-called "missing diameter," fail to fit the data in the range of sizes below ~ 800 nm, where the optical bias is most nonlinear. Consequently, a manifestly "liner fit" would predict an incorrect (and size-dependent) value of the specific conductance $G_0$ and, hence, $J_c$.

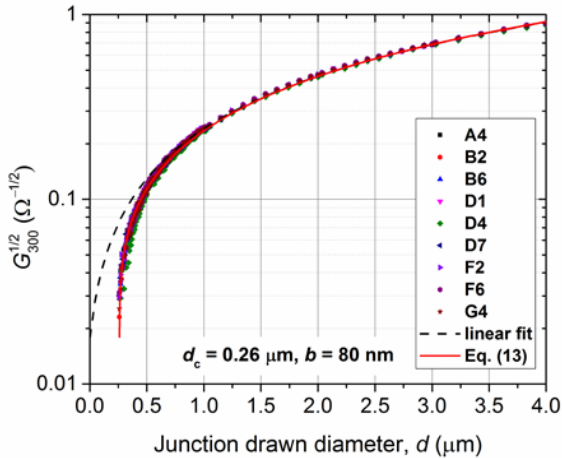

Fig. 8. Typical data showing the square root of JJ conductance at room temperature, $G_{300}^{1/2}$, as a function of the drawn diameter of JJs on the photomask. The data were taken on 18 chips at 9 different locations on the typical wafer from 10 kA/cm² process run. The locations, labeled as A4, B2, etc., correspond to the 7x7 exposure grid (A,B,…,G;1,2,…,7) of the wafer. 160 junctions with drawn diameters from 0.20 μm to 15 μm were measured at each location. The data for 1296 JJs are shown for the range of sizes from 250 nm to 4 μm. The solid line is the fit to (13) with $d_w$ given by (1). The dashed line (linear fit) is a fit to $G_{300} = G_0(\pi/4)(d - d_0)^2$ which does not account for $d_c$, where $d_0$ is the so-called "missing diameter," giving $d_0 = 80$ nm. The full fit to (13) with (1) gives $G_0 = 64$ mS/μm², $d_c = 0.26$ μm, and $b = 80$ nm.

Typical histograms of the conductance of nominally identical junctions at room-$T$ are shown in Fig. 9 for all JJs tested on a wafer. The distributions are fitted well by a Gaussian distribution with a standard deviation $\sigma_G$ that depends on the junction size. The data in Fig. 9 represent aggregate distributions of JJs located on 9 chips over 200-nm wafers, and so the $\sigma_G$ values represent wafer-level spreads of JJ conductance. Local, on-chip spreads, which are more important for SFQ circuits, are in fact smaller as will be discussed below. The wafer spread is a convolution of several factors: local variations of JJ area and tunnel barrier transparency within a 5x5 mm² chip, global variation of the mean area caused by focus variations from one exposure field to another, and global variation of the trilayer tunnel barrier specific conductance on the wafer scale.

We have explained above that the distribution of the critical currents, the relative standard deviation $\sigma I_c$, is expected to be dominated by the distribution of the JJ conductance, characterized by $\sigma_G$. To verify this, we measured the $I$-$V$ characteristics of series arrays of 1000 JJs, Fig. 10, and determined the distribution of the switching currents of JJs in the arrays, Fig. 11.

In general, the distribution of switching currents $I_s$ in the array of JJs is a convolution of the switching distribution of a single JJ [49] and the distribution of the $I_c$ of all JJs in the array. At sufficiently large $I_c$, i.e., for $I_c \gg 2\pi k_B T_{eff}/\Phi_0$, where $\Phi_0$ is the flux quantum and $T_{eff}$ is the effective noise temperature seen by the junction, the switching current $I_s$ is very close to the actual $I_c$. In this limit, the $I_s$ distribution is practically a good estimate for the $I_c$ distribution.

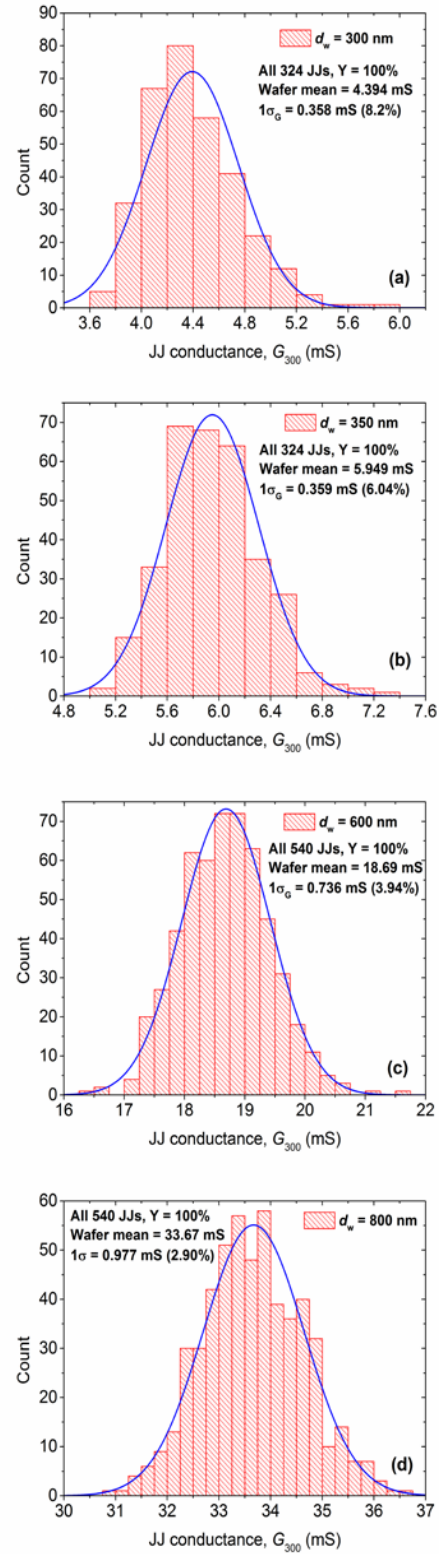

Fig. 9. The aggregate distributions of JJ conductance at room-$T$ for several realized sizes of JJs on wafer, $d_w$: (a) – 300 nm, (b) – 350 nm, (c) – 600 nm, (d) – 800 nm. The distributions represent 324 JJs measured at 9 locations (36 JJs per chip) for (a) and (b), and 540 JJs from the same locations (60 JJs per chip) for (c) and (d). Solid curves are Gaussian fits with the standard deviation $\sigma_G$ shown in each figure. JJ conductance at room-T is proportional to the normal-state conductance at 4.2 K and to $I_c$ of JJs. Therefore, the standard deviations shown here represent the standard deviation of the $I_c$ of JJs on the wafer scale. JJ yield, $Y$, on these test structures was 100%.

Conversely, for JJs with small $I_c$, the switching current can be significantly affected by thermal and external noise, leading to an $I_s$ distribution wider than the $I_c$ distribution if there is no sympathetic switching. Therefore, the width of the $I_s$ distribution for an array of JJs should be viewed as an upper boundary. We have found good agreement between the on-chip variations of JJ conductance at room-$T$, characterized by standard deviation $\sigma_G$, and the on-chip variations of the switching current of JJs characterized by standard deviation $\sigma_{Is}$ for JJ drawn sizes above ~ 700 nm. For smaller sizes, down to the smallest size of 300 nm measured, the switching distributions at 4.2 K get progressively wider than the conductance distribution for the same size of JJs, as expected.

In the following discussions, we will use the conductance distribution as an estimator for the distribution of JJ critical currents both on-chip and on-wafer. We furthermore will use relative standard deviations expressed as a percent of the mean value.

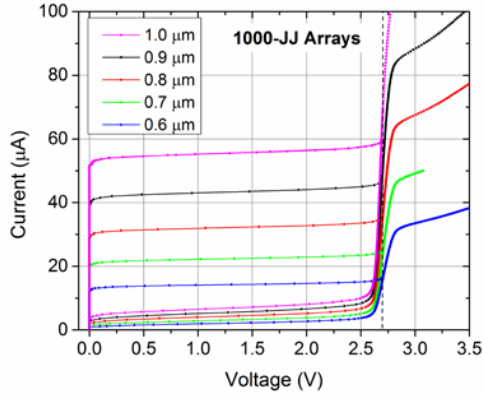

Fig. 10. The *I-V* characteristics of 1000-JJ series arrays of Josephson junctions with drawn diameters from 1.0 μm to 0.6 μm (from top to bottom). Up to 40k JJs per chip were measured using these arrays. The *I-V* curves were used to build the switching distributions shown in Fig. 11 and characterize the on-chip distributions of the critical current of JJs, $I_c$.

The conductance distribution standard deviation depends on the size of the junctions. This dependence is shown in Fig. 12 for the on-chip and wafer-averaged conductance standard deviation. If we assume that all variations of JJ conductance arise solely from variations of the JJ area and take the barrier to be absolutely uniform, we can develop a simple model to describe the data. The JJ area is $A = (\pi/4)d_w^2$, and its fluctuations are twice the fluctuations of JJ diameter (to leading order in the fluctuating term). If we assume that the area fluctuations come from fluctuations of the JJ area on the chrome photomask, so-called mask errors, we obtain using (1)

$$\frac{\partial A}{A} = \frac{2(\delta d_w)}{d_w} = \frac{2d(\delta d)}{(d^2 - d_c^2)^{1/2}[(d^2 - d_c^2)^{\frac{1}{2}} + b]}, \quad (14)$$

where $\delta d$ is the variation of the JJ diameter on the chrome photomask introduced by the photomask fabrication process. Since the address size used to e-beam-write the junction photomask is 5 nm, it is reasonable to assume $\delta d$ to be approximately this value and treat it as a fitting parameter in fitting (14) to the data in Fig. 12.

In order to confirm this model, we measured the on-chip variations of the JJ diameters at different steps of JJ fabrication, using a Hitachi CD-SEM as described previously.

Unfortunately, the accuracy of the SEM measurements suffers at JJ sizes larger than about 1 μm, because at the relatively low magnification needed to be used to image the JJ, nm-scale variations become undetectable. In contrast, 200k magnification was used for the smallest JJ sizes.

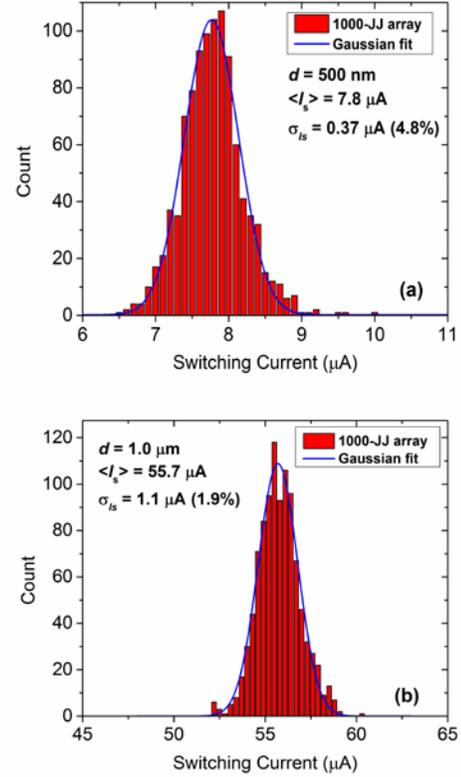

Fig. 11. The typical switching current distributions obtained from the *I-V* characteristics of 1000-JJ series arrays shown in Fig. 10 and similar: (a) – $d = 500$ nm, (b) - $d = 1000$ nm. For the critical currents much larger than the thermal noise, these distributions characterize the $I_c$ distributions.

The measured relative standard deviation of the diameter, $\sigma_d/\langle d \rangle$, where $\langle d \rangle$ is the mean diameter, was converted into the relative standard deviation of the area $\sigma_A/\langle A \rangle$. The results shown in Fig. 13 are for a typical wafer from the 8M process run. The measured area variation can be explained by the model (14) at $\delta d \approx 4$ nm. This value is very close to the grid size of 5 nm used to define the polygons approximating the circular JJs, corresponding to the address size used for e-beam writing of the photomask patterns. While this value is consistent with a "mask error" model, it does not account explicitly for any imperfections introduced during the etch process. It is worth noting that we did not observe any significant difference in the area variations of JJ features after different processing steps, except perhaps for very small JJs.

The best fit to the JJ conductance variations gives a 50% larger $\delta d \approx 6$ nm, see the solid lines in Fig. 12. Although the range of mask errors is small and likely independent of the JJ mask size, the strong enhancement of the area fluctuations observed near the minimum printable size is due to the strongly nonlinear relation between the drawn and realized sizes (1), see in Sec. 2B. This enhancement makes printing of sizes near the cut-off size $d_c$ impractical if a precise area control is required.

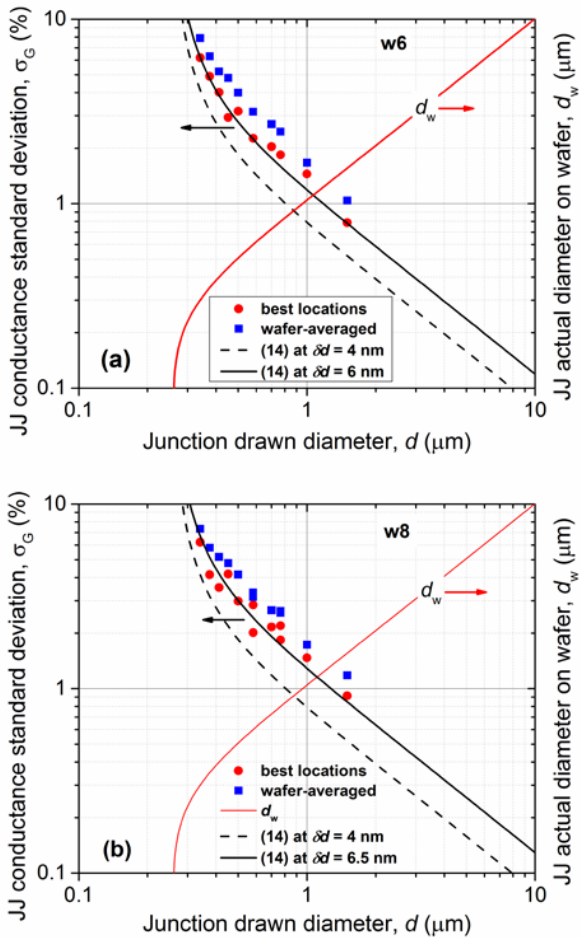

Fig. 12. The relative standard deviation of JJ conductance, in percent of the mean, as the function of JJ drawn diameter on the photomask for the two typical wafers. The actual diameter of the JJs on the wafers is shown on the right axis. The data were obtained by measuring 9 chips per wafer and the minimum of 36 to 60 JJs of the same size per chip. The wafer-averaged data are shown by squares, whereas the data from the best locations on the wafer are shown as dots. The expected conductance variations following from the model of JJ area variations (14) at $\delta d = 4$ nm as measured in SEM, see Fig. 13, is shown by the dashed curve. The best fits to the best data are shown by the solid curves and give $\delta d = 6$ nm and $\delta d = 6.5$ nm for w6 (a) and wafer 8 (b), respectively.

The JJ sizes that can be implemented in VLSI SFQ circuits with $J_c = 10$ kA/cm² range from 700 nm to 2000 nm, giving the $I_c$ range from 38 µA to 314 µA. Smaller critical current (*i.e.*, smaller sizes) would give unacceptable bit error rates in logic cells, and this determines the lower bound. For this range, the on-chip spreads of (14) at $\delta d = 4$ nm conductance and $I_s$ are less than 2%, see Fig. 12, and the on-wafer spread is less than 3%, see Fig. 9. We believe that these spreads are sufficiently low for yielding SFQ circuits with $10^6$ JJs and beyond, and they meet the requirements specified in [1] and [46].

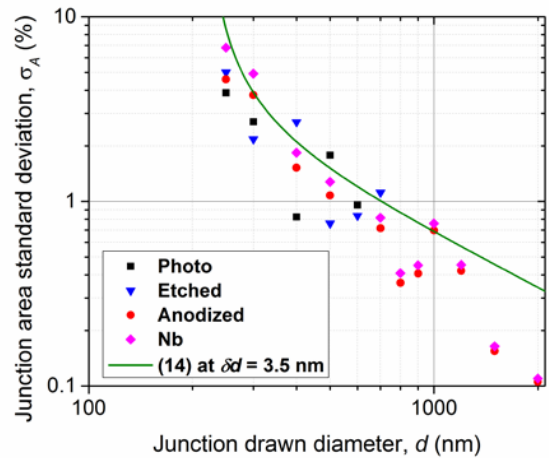

Fig. 13. The relative standard deviation of JJ feature area variations at different steps of JJs fabrication: photoresist mask, etched counter electrode, anodized counter electrode, and un-anodized part of niobium counter electrode. The diameters of multiple JJs were measured by CD-SEM to calculate the standard deviation. The accuracy of the measurements decreases at large sizes because progressively lower magnification needs to be used to image the features. Equation (14) at $\delta d = 3.5$ nm is shown by a solid curve.

## C. 20 kA/cm² process

Further increasing the speed of SFQ circuits may require $J_c$ higher than 10 kA/cm² (the present target of our 8M and 10M processes) and correspondingly smaller JJ sizes to maintain a fixed $I_c$. To make a preliminary evaluation of the expected parameter spreads, we made a process-development run targeting $J_c$ of 20 kA/cm² and 50 kA/cm². We used the same set of photomasks and the same processing as was used for the 10 kA/cm² process presented above.

The *I-V* characteristics of individual JJs and 1000-JJ arrays were measured at 4.2 K and are shown in Fig. 14. The switching distributions were extracted from the *I-V*s, and the room-*T* measurements on multiple individual junctions were done on the same set of JJs as was used in Sec. 3B.

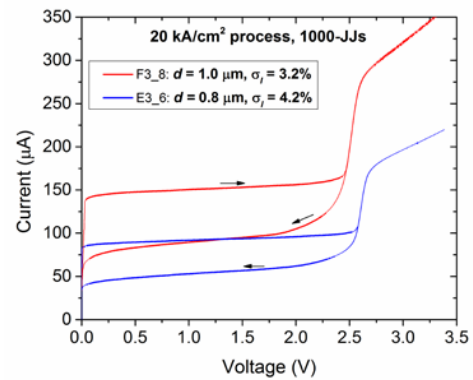

Fig. 14. The *I-V* characteristics of 1000-JJ arrays of Josephson junctions with drawn sizes 1.0 µm and 0.8 µm, from top to bottom. The exact value of $J_c$ on this wafer was 23.8 kA/cm² determined from the scaling of $I_s$ in the range of drawn sizes from 0.6 µm to 2.0 µm. The specific normal state resistance was found to be $R_{N0} = 7.57$ Ω·µm² and $J_c R_{N0} = 1.80$ mV.

## D. 50 kA/cm² process

The same fabrication and electrical measurements were done for a 50 kA/cm² process run. The typical *I-V* curves for single JJs are shown in Fig. 15.

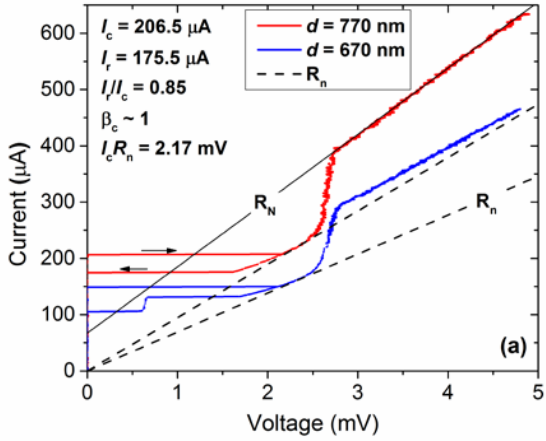

Fig. 15. The *I*-*V* characteristics of junctions with $J_c = 45$ kA/cm² and drawn diameters 770 nm and 670 nm, respectively, for the top and bottom curves. At this $J_c$, the junctions are self-shunted with damping parameter $\beta_c$ close to 1 as indicated by the return current $I_r = 0.85I_c$. Dashed lines show the subgap resistance, $R_n$ damping the JJs at these voltages. The solid line shows the normal state resistance above the gap voltage, $R_N$ for $d = 770$ nm. A current step on the return branch at $V \sim 0.65$ mV is a geometric resonance of Josephson oscillations in the JJ leads. The normal-state specific resistance, $R_{N0}$, at this $J_c$ is 4.45 $\Omega$ μm² and $J_cR_{N0} = 2.01$ mV.

An "excess" current can be seen in *I*-*V* curves defined as the zero voltage intercept of the solid line in Fig. 15. It is about $0.33I_c$. This "excess" current results from the time averaging of Josephson oscillations above the gap voltage of the current-biased JJs [34], [35].

The advantage of the 50 kA/cm² process is that JJs at this current density intrinsically become almost nonhysteretic as can be seen in Fig. 15. The return current is about 85% of the critical current, indicating self-shunting with the damping parameter $1 < \beta_c < 2$. Therefore, the junctions can be used in SFQ circuits without external resistive shunts. Their damping is provided by the internal subgap resistance $R_n$ and the resultant $I_cR_n$ product is about 2 mV, a factor of 2 higher than for the externally shunted JJs in the 10 kA/cm² process at the critical damping $\beta_c = 1$. Such junctions would enable much faster and denser circuits at future technology nodes.

The JJ conductance spreads and $I_s$ spreads for JJs with 24 kA/cm² and 45 kA/cm² are shown in Fig. 16. They are somewhat larger than the typical spreads shown in Sec. 3B for the 10 kA/cm² process, but follow similar trends.

The spreads can be fitted by (14) with $\delta d = 10.5$ nm. Since we used the same photomasks and fabrication process, this difference may indicate the existence of another factor that contributes to the spreads of the effective, electrical, area of the JJs in addition to the geometric fluctuations described by (14) and that scales with JJ diameter in the same manner. Although the nature of a higher variability of high-$J_c$ JJs has not yet been firmly established, we anticipate that these initial spreads can be reduced to the same level as those for the 10 kA/cm² with additional fabrication runs and process maturity.

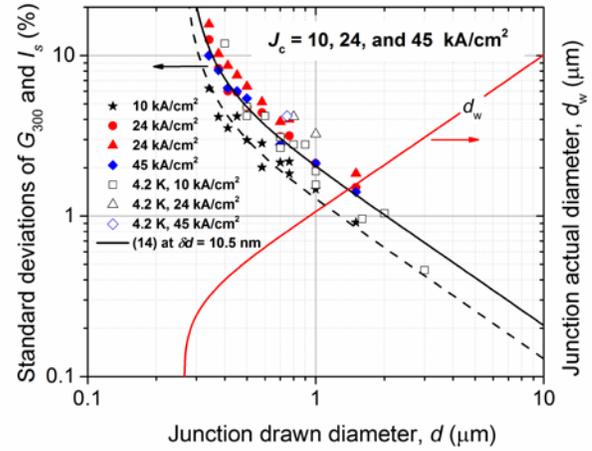

Fig. 16. The standard deviation of JJ conductance at room-*T* for single JJs and of the switching current at 4.2 K for 1000-JJ arrays fabricated with critical current densities $J_c = 10$ kA/cm², 24 kA/cm², and 45 kA/cm². The data from top to bottom correspond to the best locations on wafers at $J_c = 10$ kA/cm² and 24 kA/cm², wafer-averaged data for 24 kA/cm², best locations at 45 kA/cm² and 4.2 K data for arrays after that. The solid line shows the expected standard deviations from model (14) at $\delta d = 10.5$ nm and fits the data for high-$J_c$ junctions very well. The fit for the 10 kA/cm² process is shown by the dashed line ($\delta d = 6$ nm), see also Fig. 12.

### E. Discussion

In the prior discussion we left aside other possible contributions to JJ variability beyond the enhanced mask errors described by (14). The first is the variation of the cut-off size $d_c$ caused by variations of focus, local photoresist chemistry, *etc*. This contribution is also enhanced by the strongly nonlinear projection present near the cut-off. Indeed, differentiating (1) with respect to $d_c$ we get a similarly diverging expression

$$\frac{|\partial A|}{A} = \frac{2d_c(\delta d_c)}{(d^2 - d_c^2)^{1/2}[(d^2 - d_c^2)^{\frac{1}{2}} + b]}, \quad (15)$$

where $\delta d_c$ is the variation of the cut-off size. The contribution of (15) to on-chip variability should be much smaller than (14), due to the much smaller numerator - $\delta d_c$ within the die should be very small. Nonetheless, it may contribute to cross-wafer variation, due to focus variation between stepped exposure fields, which could produce $\delta d_c$ in the range of $\pm 10$ nm across the wafer.

The second is variations of electric biases $d_0$ and $d_1$ in (13), which is equivalent to variations of $b$ in (1)

$$\frac{\partial A}{A} = \frac{2(\delta b)}{[(d^2 - d_c^2)^{\frac{1}{2}} + b]}. \quad (16)$$

These $\delta b$ variations could be caused by contamination of Nb around JJ sidewalls, *e.g.*, hydrogen absorption during etching [36]-[39], that affects superconducting properties of Nb near the Nb/Nb$_2$O$_5$ interface. The size of this potential variability, or if it even exists, is not known. The total area variance is the sum of all variances described by (14)-(16).

Yet another potential source of JJ variability is fluctuations of the barrier transparency, especially in high-$J_c$ junctions. The possibility of the existence of nanoscale regions with high barrier transparency, often called pinholes or quantum point contacts (QPCs), in ultrathin $AlO_x$ tunnel barriers was assumed from the early days of transport measurements on tunnel junctions, see [52] and references therein. It was also invoked to explain the details of the subgap transport, see, *e.g.*, [7], [8], [35], [50]-[54]. Recently, wide distributions of $AlO_x$ barrier thickness in $Al/AlO_x/Al$ junctions were found using scanning transmission electron microscopy [55],[56], confirming the existence of nanoscale regions with much thinner barrier than the mean thickness.

The existence of these QPCs does not necessarily imply a higher variability of high-$J_c$ junctions. It depends on their spatial and transparency distribution functions. For instance, for junctions effectively saturated with QPCs channels, if the distribution of QPC transparencies is very broad and universal [57], as was assumed in [35] and [58], the distribution-averaged conductance and $I_c$ should be self-averaging quantities that do not vary significantly from junction to junction. On the other hand, if the QPCs transparency distribution is quite narrow and fabrication-dependent [50], [54], such that the transport properties are governed by QPCs with the transparency close to 1, the number of these QPCs, $n$, in a high-$J_c$ junction is expected to be on the order of a few hundred [51], [53], and scale proportionally to the junction area. Statistical fluctuations in the number of high-transparency QPCs are expected to be on the order of $n^{-1/2}$ and scale as $A^{-1/2}$. This would result in $\sigma_G$ dependence similar to (16) with a quantity in the numerator characterizing the mean spacing between the QPCs. We leave a more detailed investigation of these issues for a separate study.

As we have shown above, the on-chip and cross-wafer spreads that we have achieved in our 10 kA/cm$^2$ processes with 4 and 8 metal layers are at a sufficient level to enable the fabrication of VLSI superconducting circuits with 10$^6$ JJs and beyond. The spreads reported here are significantly lower than those reported for the best industrial [41]-[45], or advanced research processes [17],[19],[40]. However, further lowering of JJ spreads is always desirable, especially for higher-$J_c$ processes with smaller JJs. The results obtained in this work show three clear tactics to accomplish this.

The first tactic is the reduction of "mask errors" – fluctuations of JJ area on the photomask – by implementing a smaller grid size for JJ design and a smaller address size for photomask e-beam writing. We are planning to utilize photomasks with a 1-nm design grid and assess improvements in future process runs. This should also allow us to verify our model (14).

The second tactic is the minimization of mask error enhancement caused by the proximity of JJ sizes to print to the photolithography cut-off size $d_c$ at which the area fluctuations diverge. This can be done by transitioning to 193-nm photolithography, which significantly reduces $d_c$. Indeed, using (9), we can estimate the cut-off size for our 193-nm exposure tool with $NA = 0.75$ to be 110 nm for the threshold resist and coherent illumination, and about 130 nm for actual resists. This is a factor of 1.6 - 2 less than for 248-nm photolithography used in this work. Then, the ratio of the minimum JJ size we need to print $d_{min} \sim 500$ nm, *e.g.*, as could be used in a 50 kA/cm$^2$ process, to the cut-off size $d_c$ will increase from the present $d_{min}/d_c \sim 2$ to $d_{min}/d_c \sim 4$, moving the region of nonlinear projection – *i.e.*, the strongly nonlinear part of the $d_w(d)$ dependence and the associated mask error enhancement - farther away from the size range of interest.

The third tactic is the identification and elimination of the extra JJ variability that accounts for approximately a 50% increase of the conductance spreads with respect to the geometric area fluctuations described by (14), see Fig. 12 and Fig. 16. We leave the implementation of these approaches for future work.

## IV. Conclusion

We have developed a fabrication process for fully planarized Nb/Al-$AlO_x$/Nb Josephson junctions on a 200-mm-wafer tool set available in a CMOS foundry with 248-nm photolithography. We have measured the on-chip and cross-wafer spreads of the junction conductance and critical current for junction sizes ranging from 200 nm to 1.5 µm in diameter and critical current densities up to 45 kA/cm$^2$. We have determined explicitly the relationship between the drawn junction size, the size realized on the wafer, and the minimum printable size for circular-shaped junctions. We have shown that a model accounting for the enhancement of mask errors near the minimum printable size accurately describes the variations of JJ parameters as a function of the drawn size. For the range of JJ sizes intended for use in SFQ circuits, we have achieved levels of JJ repeatability and yield consistent with the large-scale integration of superconducting circuits comprising 10$^6$ JJs and beyond. The junctions and 8-metal-layer fully planarized fabrication process described in this work have been utilized to fabricate with high yield and successfully demonstrate RQL shift registers with 72800 JJs per 5 mm by 5 mm chip [59]. A truncated, 4-metal-layer, version of this process has been used to successfully demonstrate a new type of ac-powered SFQ circuits with 32800 Josephson junctions [60]. To the best of our knowledge, these are the largest SFQ circuits demonstrated up to date.

## Acknowledgment

We greatly appreciate discussions with Dr. Marc Manheimer and Dr. D. Scott Holmes as well as their interest and support of this work. We would like to thank George Fitch for his help in CAD of Josephson junctions, test structures, and photomasks, and Jeffrey Knecht for his help with PROLITH simulator and discussions of 193-nm photolithography.

## References

[1] *Superconducting Technology Assessment*, National Security Agency, Office of Corporate Assessments, August 2005. [Online]. Available: http://www.nitrd.gov/pubs/nsa/sta.pdf

[2] S. Anders, M.G. Blamire, F.-Im. Buchholz *et al.* "European roadmap on superconductive electronics – status and perspectives," *Physica C*, vol. 470, pp. 2079-2126, 2010.

[3] *100 Years of Superconductivity*, eds. H. Rogalla and P.H. Kes, 2012 (Taylor and Francis, Boca Raton, FL, USA).

[4] M.B. Ketchen, D. Pearson, A.W. Kleinsasser, *et al*. "Sub-µm, planarized Nb-AlO$_x$-Nb Josephson process for 125 mm wafers developed in partnership with IBM si technology," *Appl. Phys. Lett*., vol. 59, pp. 2609-2611, Nov. 1991.

[5] Z. Bao, M. Bhushan, S. Han, and J.E. Lukens, "Fabrication of high quality, deep-submicron Nb/AlOx/Nb Josephson junctions using chemical mechanical polishing," *IEEE Trans. Appl. Supercond.* vol. 5, pp. 2731-2734, 1995.

[6] D.J. Flees, "Experimental studies of band-structure properties in Block transistors," Ph.D. Dissertation, Dept. of Physics and Astronomy, State University of New York at Stony Brook, Stony Brook, NY, 1998.

[7] A.W. Kliensasser, R.E. Miller, W.H. Mallison, and G.D. Arnold, "Observation of multiple Andreev reflections in superconducting tunnel junctions," *Phys. Rev. Lett*. vol. 72, pp. 1738-1741, 1994.

[8] V. Patel and J.E. Lukens, "Self-shunted Nb/AlOx/Nb Josephson junctions," *IEEE Trans. Appl. Supercond*. vol. 9, pp. 3247-3250, 1999.

[9] K.K. Berggren, E. Macedo, D.A. Feld, J.P. Sage, "Low $T_c$ superconductive circuits fabricated on 150-mm-diamter wafers using a doubly planarized Nb/AlOx/Nb process," *IEEE Trans Appl. Supercond.,* vol. 9, pp. 3271-3273, 1998.

[10] B. Bumble, A. Fung, A.B. Kaul, A.W. Kleinsasser, G.L. Kerber, P. Bunyk, and E. Ladizinsky, "Submicrometer Nb/Al-AlOx/Nb integrated circuit fabrication process for quantum computing applications," *IEEE Trans. Appl. Supercond.*, vol. 19, pp. 226–229, Jun. 2009.

[11] S.K. Tolpygo, D.J.C. Amparo, R.T. Hunt, J.A. Vivalda, and D.T. Yohannes, "Subgap leakage in Nb/Al-AlOx/Nb Josephson junctions and run-to-run reproducibility: Effects of oxidation chamber and film stress," *IEEE Trans. Appl. Supercond*., vol. 23, p. 1100305, June 2013.

[12] W. Chen, A.V. Rylyakov, V. Patel, J.E. Lukens, and K.K. Likharev, "Rapid single flux quantum T-flip-flop operating at 770 GHz," *IEEE Trans. Appl. Supercond.*, vol. 9, pp. 3212–3215, June 1999

[13] S.K. Tolpygo, V. Bolkhovsky, T. Weir, L. Johnson, W.D. Oliver, and M.A. Gouker, "MIT LL superconductor electronics fabrication process for VLSI circuits with 4, 8, and 10 niobium layers," *Superconductivity New Forum (SNF) Global Edition*, Issue No. 30, Preview No. 2, Art. ID. STP402, Sept. 2014. [On-line]. Available: http://snf.ieeecsc.org/issue-30-october-2014-preview-number-2. *Appl. Supercond. Conf., ASC* 2014, Charlotte, NC, 11-15 August 2014, paper 1EOr2A-01; unpublished.

[14] S.K. Tolpygo, V. Bolkhovsky, T.J. Weir, C.J. Galbraith, L.M. Johnson, M.A. Gouker, and V.K. Semenov, "Inductance of circuit structures for MIT LL superconductor electronics fabrication process with 8 niobium layers," *arXiv*: 1408.5828. [On-line]. Available: http://arxiv.org/abs/1408.5828. *IEEE Trans. Appl. Supercond.*, vol. 25, no. 3, part 1, pp. ? 2015. Doi: 10.1109/TASC.2014.2369312.

[15] S.K. Tolpygo, V. Bolkhovsky, T. Weir, L. Johnson, W.D Oliver, and M.A. Gouker, "Deep sub-micron stud-via technology of superconductor VLSI circuits," *Supercond. Sci. Technol.*, vol. 27, 025016, Jan. 2014. doi:10.1088/0953-2048/27/2/025016

[16] R.E. Miller, W.H. Mallison, A.W. Kleinsasser, K.A. Delin, and E.M. Macedo, "Niobium trilayer Josephson tunnel junctions with ultra-high critical current densities," *Appl. Phys. Lett.*, vol. 63, pp. 1423-1425, 1993.

[17] H. Akaike, Y. Kitagawa, T. Satoh, K. Hinode, S. Nagasawa, and M. Hidaka, "Effect of photomask pattern shape for a junction counter-electrode on critical current uniformity and controllability in Nb/AlOx/Nb junctions," *IEEE Trans. Appl. Supercond.* vol. 15, pp. 102-105, 2005.

[18] S. Nagasawa, T. Satoh, K. Hinode, Y. Kitagawa, M. Hidaka, H. Akaike, A. Fujimaki, K. Takagi, N. Takagi, N. Yoshikawa, "New Nb multi-layer fabrication process for large-scale SFQ circuits," *Physica C*, vol. 469, pp. 1578-1584, 2009.

[19] S. Nagasawa, K. Hinode, T. Satoh, M. Hidaka, H. Akaike, A. Fujimaki, N. Yoshikawa, K. Takagi, and N. Takagi, "Nb 9-layer fabrication process for superconducting large-scale SFQ circuits and its process evaluation," *IEICE Trans. Electron*., vol. E97-C, pp. 132-140, Mar. 2014.

[20] A. K.-K. Wong, *Optical Imaging in Projection Microlithography*, Bellingham, WA, USA: SPIE Press, 1985, p. 83.

[21] J.W. Goodman, *Introduction to Fourier Optics*, 2nd ed., Boston, MA, USA: The McGraw-Hill, 1996.

[22] M. Born and E. Wolf, *Principles of Optics*, 6$^{th}$ ed., Oxford, U.K.: Pergamon Press, 1980, pp. 378-401.

[23] A. K.-K. Wong, *Resolution Enhancement Techniques*, Bellingham, WA, USA: SPIE Press, 2001, p. 34.

[24] C. Mack, *Fundamental Principles of Optical Lithography: The Science of Microfabrication*, Chichester, England: John Wiley & Sons, 2007, p. 122.

[25] F.M. Schellenberg and C. Mack, "MEEF in theory and practice," *Proc. SPIE*, vol. 3873, 19th Annual Symposium on Photomask Technology, pp. 189-202, Dec. 1999. doi:10.1117/12.373313. Available: http://dx.doi.org/10.1117/12.373313

[26] PROLITH 9.0 Lithography Simulation Tool, KLA-Tencor. [Online]. Available: http://www.kla-tencor.com/lithography-modeling/chip-prolith.html

[27] G.L. Kerber, L.A. Abelson, K. Edwards, R. Hu, M.W. Johnson, M.L. Leung, and J. Luine, "Fabrication of high current density Nb integrated circuits using a self-aligned junction anodization process," *IEEE Trans. Appl. Supercond.*, vol. 13, pp. 82-86, June 2003.

[28] X. Meng and T. Van Duzer, "Light-anodization process for high-$J_c$ micron and submicron superconducting junction and integrated circuit fabrication," *IEEE Trans. Appl. Supercond.*, vol. 13, pp. 91–94, Jun. 2003.

[29] D. Nakada, K.K. Berggren, E. Macedo, V. Liberman, and T. P. Orlando, "Improved critical-current-density uniformity by using anodization," *IEEE Trans. Appl. Supercond.*, vol. 13, pp. 111–114, Jun. 2003.

[30] J.G. Simmons, "Generalized formula for the electric tunnel effect between similar electrodes separated by a thin insulating film," *J. Appl. Phys.*, vol. 34, pp. 1793–1803, Jun. 1963.

[31] A. Kleinsasser, T. Chui, B. Bumble, and E. Ladizinsky, "Critical current density and temperature dependence of Nb-Al Oxide-Nb junction resistance and implications for room temperature characterization," *IEEE. Trans. Appl. Supercond*., vol. 23, p. 1100405, Jun. 2013.

[32] K.K. Berggren, M. O'Hara, J. P. Sage, and A.H. Worsham, "Evaluation of critical current density of Nb/Al/AlOx/Nb Josephson junctions using test structures at 300 K," *IEEE Trans. Appl. Supercond.*, vol. 9, pp. 3236–3239, Jun. 1999.

[33] V. Ambegaokar and A. Baratoff, "Tunneling between superconductors," *Phys. Rev. Lett.*, vol. 10, pp. 486–489, Jun. 1963.

[34] D. Averin and A. Bardas, "Josephson effect in a single quantum channel," *Phys. Rev. Lett*., vol. 75, pp. 1831-1834, 1995.

[35] Y. Naveh, V. Patel, D.V. Averin, K.K. Likharev, and J.E. Lukens, "Universal distribution of transparencies in highly conductive Nb/AlOx/Nb junctions," *Phys. Rev. Lett*. vol. 85, pp.5404-5407, 2000.

[36] K. Hinode, T. Satoh, S. Nagasawa, and M. Hidaka, "Hydrogen-inclusion -induced variation of critical current in Nb/AlOx/Nb Josephson junctions," *J. Appl. Phys.*, vol. 104, pp. 023909-1–023909-6, Jul. 2008.

[37] D. Amparo and S.K. Tolpygo, "Investigation of the role of H in fabrication-process-induced variations of Nb/Al/AlOx/Nb Josephson junctions," *IEEE Trans. Appl. Supercond.*, vol. 21, pp. 126-130, June 2011.

[38] S.K. Tolpygo, D. Amparo, R.T. Hunt, J.A. Vivalda, and D.Y. Yohannes, "Diffusion stop-layers for superconducting integrated circuits and qubits with Nb-based Josephson junctions," *IEEE Trans. Appl. Supercond.*, vol. 21, pp. 119-125, Jun. 2011.

[39] S.K. Tolpygo and D. Amparo, "Fabrication-process-induced variations of Nb/Al/AlOx/Nb Josephson junctions in superconductor integrated circuits," *Supercond. Sci. Technol.*, vol. 23, 034024, Feb. 2010. doi:10.1088/0953-2048/23/3/034024

[40] S. Nagasawa, K. Hinode, T. Satoh, H. Akaike, Y. Kitagawa, and M. Hidaka, "Development of advanced Nb process for SFQ circuits," *Physica C*, vol. 412, pp. 1429–1436, Oct. 2004.

[41] D. Yohannes, S. Sarwana, S.K. Tolpygo, A. Sahu, Y.A. Polyakov, and V.K. Semenov, "Characterization of HYPRES 4.5 kA/cm$^2$ and 8 kA/cm$^2$ Nb/AlOx/Nb fabrication processes," *IEEE Trans. Appl. Supercond*., vol. 15, pp. 90-93, June 2005.

[42] D. Yohannes, A. Kirichenko, S. Sarwana, and S.K. Tolpygo, "Parametric testing of HYPRES superconducting circuit fabrication process," *IEEE Trans. Appl. Supercond*., vol. 17, pp. 181-186, June 2007.

[43] S.K. Tolpygo, D. Yohannes, R.T. Hunt, J.A. Vivlada, D. Donnelly, D. Amparo, and A.F. Kirichenko, "20 kA/cm$^2$ process development for superconducting integrated circuits with 80 GHz clock frequency," *IEEE Trans. Appl. Supercond*., vol. 17, pp. 946-951, June 2007.

[44] L.A. Abelson and G.L. Kerber, "Superconductor integrated circuit fabrication technology," *Proc. IEEE*, vol. 92, no. 10, pp. 1517-1533, Oct. 2004.

[45] A.D. Smith, S.L. Thomasson, and C. Dang, "Reproducibility of niobium junction critical currents: Statistical analysis and data," *IEEE Trans. Appl. Supercond.*, vol. 3, pp. 2174–2177, Mar. 1993.

[46] P. Bunyk, K. Likharev, and D. Zinoviev, "RSFQ technology: physics and devices," *Int. J. High Speed Electron. Syst.*, vol. 11, pp. 257-305, 2001.

[47] *Physics and Applications of Mesoscopic Josephson Junctions*, eds. H. Ohta and C. Ishii, Tokyo, Japan: The Physical Society of Japan, 1999.

[48] L.A. Goodman, "On the exact variance of the products," *J. American Stat. Assn.*, vol. 55, pp. 708-713, 1960.

[49] T. Fulton and L.N. Dukleberger, "Lifetime of the zero-voltage state in Josephson tunnel junctions," *Phys. Rev. B*, vol. 9, no. 11, pp. 4760-4768, June 1974.

[50] W.H. Mallison, R.E. Miller, and A.W. Kleinsasser, "Effect of growth conditions on the electrical properties of Nb/Al-oxide/Nb tunnel junctions, "*IEEE Trans. Appl. Supercond.*, vol. 5, no. 2, pp. 2330-2332, June 1995.

[51] A.W. Kleinsasser, "High performance Nb Josephson devices for petaflops computing," *IEEE Trans. Appl. Supercond.*, vol. 11, no. 1, pp. 1043-1049, Mar. 2001.

[52] J.M. Rowell and W.L. Feldman, "Excess currents in superconducting tunnel junctions," *Phys. Rev.*, vol. 172, no. 2, pp. 393-401, Aug. 1968.

[53] V. Patel, S.K. Tolpygo, W. Chen, and J.E. Lukens, "Fabrication and properties of Nb/AlOx/Nb self-shunted Josephson junctions with high critical current densities," *Extended Abstracts of 7th International Superconductive Electronics Conference, ISEC'99*, PI5.1, pp. 229- 231, Berkeley, CA, June 21-25, 1999.

[54] S.K. Tolpygo and D. Amparo, "Electric stress effect on Josephson tunneling through ultrathin AlOx barrier in Nb/Al/AlOx/Nb junctions," *J. Appl. Phys.*, vol.104, 063904, 2008; doi: 10.1063/1.2977725.

[55] T. Aref, A. Averin, S. van Dijken, A. Ferring, M. Koberidze, V.F. Maisi, H.Q. Nguyend, R.M. Nieminen, J.P. Pekola, and L.D. Yao, "Characterization of aluminum oxide tunnel barriers by combining transport measurements and transmission electron microscopy imaging," *J. Appl. Phys.*, vol. 116, no.7, pp. 073702,073702-4, Aug. 2014. Doi.: 10.1063/1.4893473

[56] L.J. Zeng, S. Nik, T. Greibe, C.M. Wilson, P. Delsing, E. Olsson, "Direct observation of the thickness distribution of ultra thin AlOx barrier in Al/AlOx/Al Josephson junctions," *arXiv*: 1407.0173v1. On-line: http://arxiv.org/abs/1407.0173

[57] K.M. Schep and G.E.W. Bauer, "Universality of transport through dirty interfaces," *Phys. Rev. Lett.*, vol. 78, no. 15, pp. 3015-3018, Apr. 1997.

[58] V. Lacquaniti, M. Belogolovskii, C. Cassiago, N. De Leo, M. Fretto, and A. Sosso, "Universality of transport properties of ultra-thin oxide films," *New J. Phys.*, vol. 14, 023025, Feb. 2012. doi:10.1088/1367-2630/14/2/023025

[59] M. Stoutimore, Q.P. Herr, A. Herr; H. Hearne, O. Oberg, J. Strong, "Reciprocal quantum logic circuits fabricated with sub-micron features," *Applied Superconductivity Conference, ASC 2014*, Charlotte, NC, USA, Aug. 10-15, 2014, presentation 4EOr3A-06, unpublished

[60] V.K. Semenov, Y.A. Polyakov, and S.K. Tolpygo, "New AC powered SFQ digital circuits," *IEEE Trans. Appl. Supercond.*, submitted for publication, ASC2014 paper 4EOr3A-01.

[61] S.J. Proctor and L.W. Linholm, "A direct measurements of interfacial contact resistance," *IEEE Electron. Dev. Lett.*, vol. EDL-3, pp. 294-296, Oct. 1982.

[62] T.A. Scheyer and K.C. Saraswat, "A two-dimensional analytical model of the Cross-Bridge Kelvin Resistor," *IEEE.Electron. Dev. Lett.*, vol. EDL-7, pp. 661-663, Dec. 1986.